\documentclass[useAMS,usenatbib,usegraphicx]{mn2e}
\usepackage{graphicx}
\usepackage{amssymb}
\usepackage{color}

\title[Photometric study of the Blazhko star {MW~Lyr} II.]{An extensive photometric study of the Blazhko RR Lyrae star MW~Lyr:
II. Changes in the physical parameters\thanks{Based on observations collected mainly with the automatic 60 cm telescope of Konkoly Observatory, Budapest, Sv\'abhegy}}
\author[J. Jurcsik et al.]{J. Jurcsik$^{1}$\thanks{E-mail: jurcsik@konkoly.hu},
\'A. S\'odor$^{1}$, B. Szeidl$^{1}$, Z. Koll\'ath$^{1}$, H. A. Smith$^{4}$, Zs. Hurta$^{1,2}$, \and M. V\'aradi$^{1,3}$, A. Henden$^{5}$, I. D\'ek\'any$^{1}$, I. Nagy$^{2}$, K. Posztob\'anyi$^{6}$, A. Szing$^{7}$, \and K. Vida$^{1,2}$, and N. Vityi$^{2}$\\
\\
$^{1}$Konkoly Observatory of the Hungarian Academy of Sciences, H--1525 Budapest PO Box 67, Hungary\\
$^{2}$Dept. of Astronomy, E\"otv\"os University, H--1518 Budapest PO Box 49, Hungary\\
$^{3}$Observatoire de Gen\'eve, Universite de Gen\`eve, CH--1290, Sauverny, Switzerland\\
$^{4}$Dept. of Physics and Astronomy, Michigan State Univ., East Lansing, MI 48824, USA\\
$^{5}$American Association of Variable Star Observers, 49 Bay State Road, Cambridge, MA 02138, USA\\
$^{6}$AEKI, KFKI Atomic Energy Research Institute, Thermohydraulic Department, H--1525 Budapest 114, PO Box 49, Hungary\\
$^{7}$University of Szeged, Dept. of Exp. Physics and Astron. Obs., H--6720 Szeged, D\'om t\'er 9, Hungary
}
\begin{document}

\date{Accepted 2008 ..... Received 2008 ...; in original form 2008 Jul 24}

\pagerange{\pageref{firstpage}--\pageref{lastpage}} \pubyear{2008}

\maketitle

\label{firstpage}

\begin{abstract}
The analysis of the multicolour photometric observations of MW Lyr, a large modulation amplitude Blazhko variable, shows for the first time how the mean global physical parameters vary during the Blazhko cycle. About $1-2\%$ changes in the mean radius, luminosity and surface effective temperature are detected. The mean radius and temperature changes are in good accordance with pulsation model results, which show that these parameters do indeed vary within this order of magnitude if the amplitude of the pulsation changes significantly. We interpret the phase modulation of the pulsation to be a consequence of period changes. Its magnitude corresponds exactly what one expects from the detected changes of the mean radius assuming that the pulsation constant remains the same during the modulation. Our results indicate that during the modulation the pulsation remains purely radial, and the underlying mechanism is most probably a periodic perturbation of the stellar luminosity with the modulation period.
\end{abstract}


\section{Introduction}

Neither multicolour photometric nor spectroscopic observations have been previously obtained with good enough time coverage of both the pulsation and modulation cycles of a Blazhko RR Lyrae star that they could be used to derive any phenomenological conclusion about the changes in the global physical properties of the star during the modulation cycle. To cover  the pulsation variations of different shape in each phase of the Blazhko modulation with observations several hundreds of hours of observing time were needed. This can be achieved only with telescopes 'dedicated' to the study of the phenomenon. As $80-90\%$ of the telescope time of the automated 60cm telescope of the Konkoly Observatory is allocated to study RR Lyrae stars, during recent years we could first obtain $BVR_CI_C$ photometric time series of Blazhko variables that are extended enough for such an investigation.

We have detected $0.010-0.005$~mag systematic changes in the intensity weighted mean $\langle{V}\rangle, \langle{B}\rangle-\langle{V}\rangle$, and $\langle{V}\rangle-\langle{I_C}\rangle$ brightness and colours of RR Gem and SS Cnc in different phases of the Blazhko modulation (see Fig. 12 and Fig. 9 in \citet{rrgI} and \cite{sscnc}, respectively). The intensity weighted quantities indicated slight brightness and temperature increases in RR Gem at the time of the largest amplitude of the pulsation, while in SS Cnc the brightest mean magnitude and bluest mean $\langle{V}\rangle-\langle{I_C}\rangle$  colour occurred during the decreasing pulsation amplitude phase of the modulation. The small amplitude of the modulation of these stars, and also the ambiguity of the equivalent static colours of RR Lyrae stars \citep{bono} make, however, these results somewhat uncertain.

The light curve analysis, utilizing mostly the $V$ data of our $\sim 1000$ hours extended multicolour CCD observations of MW Lyr, an RR Lyrae star showing large amplitude Blazhko modulation, was presented in \citet[hereafter Paper I]{mw1}. The large amplitude of the modulation of MW Lyr makes it possible to detect changes in the mean magnitudes and colours more definitely if there are any indeed.

In order to exploit the  most information possible from the multicolour light curves of RR Lyrae stars we have recently developed the IP method \citep[Inverse Photometric method,][]{cikk}. This method gives good estimates of the mean physical parameters and their variations with pulsation phase exclusively from photometric data without any spectroscopic observations. Using the extended $BVI_C$ time series of MW Lyr the IP method shows the differences between the pulsation in different phases of the modulation, and whether or not there are any changes in the mean global physical parameters of the star during the Blazhko cycle.

\section{Data and method}

\begin{figure*}
  \includegraphics[width=17.2cm]{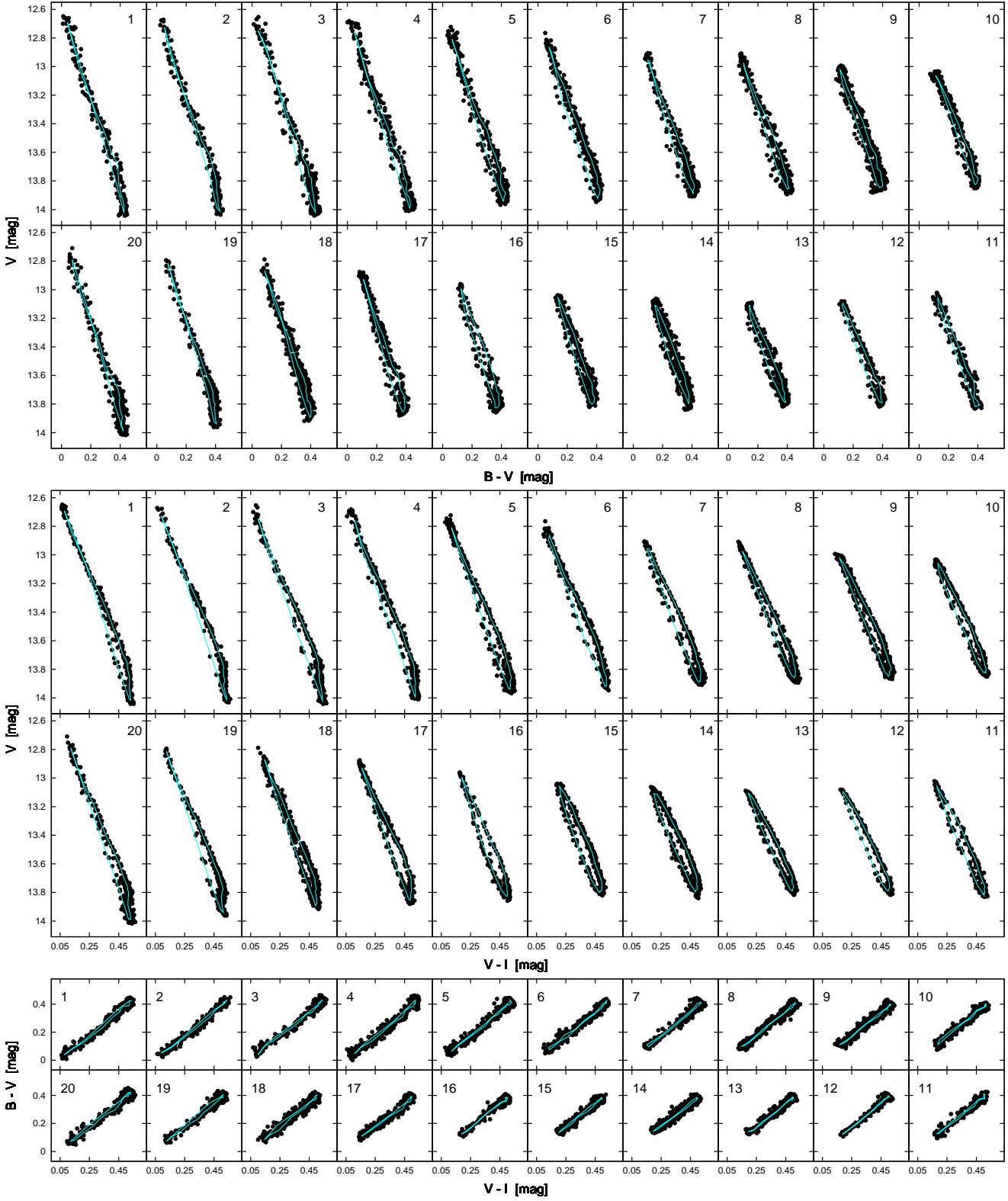}
  \caption{$V vs. B-V$,   $V vs. V-I_C$, and $B-V vs. V-I_C$ data of MW Lyr are shown in 20 different phase bins of the Blazhko modulation. The drawn curves connect the mean values of the magnitudes in 20 phase bins of the pulsation cycle.}
\label{lck}
\end{figure*}

The photometric data utilized in this paper were published in Paper~I. The observed colours and magnitudes are dereddened using $E(B-V)=0.10$~mag interstellar extinction given by the \cite{schlegel} catalogue and  $A_V=3.1E(B-V)$ and $E(V-I_C)=1.27E(B-V)$ relations. 

In the present paper the mean light and colour curves of MW Lyr and the actual light and colour curves in 20 different phase bins of the 16.546 d modulation cycle are investigated. The colour-magnitude and two-colour loops of the pulsation in different phases of the modulation are shown in Fig.~\ref{lck}. This is the first multicolour photometric observation of a Blazhko variable that makes a detailed study of the colour behaviour changes during the Blazhko cycle possible.

The analysis is performed using the IP method \citep{cikk}, that finds which  $T_{eff}$ and $V_{rad}$ variations during the pulsation results in luminosity, radius and effective gravity changes that best match the light curves using synthetic magnitudes and bolometric corrections of static atmosphere models \citep{kurucz}. The input data of the IP method are the Fourier fits of the time series data. Solutions are obtained using the Fourier representations of either the $V$, $B-V$, and $V-I_C$ or the $V, B,$ and $I_C$ time series. 

Those parameters that definitely must not change with the Blazhko period, namely the metallicity, mass and the distance of the star are determined from the mean light curves of the entire data set and are fixed when studying the changes during the Blazhko cycle. Two mean data sets are investigated to determine these parameters, one corresponding to the original observations, and another where data are corrected for the phase modulation with appropriate time transformation of the observations. In Paper I it was demonstrated that if  phase modulation is due to pulsation period changes during the Blazhko cycle, this treatment is valid and the mean light curve of the time transformed data represents the mean pulsation variation better than the mean light curve of the original data. 

As no spectroscopic observation of MW Lyrae has ever been obtained, the metallicity is calculated from the \cite{jk96} formula which derives the [Fe/H] from the period and the $\Phi_{31}$ phase difference of the $V$ light curve. This formula gives $-0.6$ and $-0.4$ [Fe/H] values for the mean light curves  of the original and the time transformed data, respectively. Therefore, it seems to be a reasonable choice to use the IP method with [M/H]=$-0.5 $ model atmosphere grid for MW Lyr assuming solar scaled chemical composition distribution as significant enhancement of the $\alpha$ elements are observed typically only in more metal poor variables \citep[see e.g., Fig 9. in][]{alpha}.

The mass and distance of MW Lyr are derived by applying the IP method using the [Fe/H]=$-0.5$ model atmosphere grid on the mean light and colour curves of the entire original and the time transformed  data sets. One basic input of the IP method is an initial $V_{rad}(\varphi)$ curve ($\varphi$ means the pulsation phase)  which is varied in the minimization process. \cite{cikk} defined this initial $V_{rad}(\varphi)$ curve twofold, first using the template $V_{rad}(\varphi)$ curve scaled according to the  $A_{V_{rad}} - V_{Amp}$ relation given by \cite{liu}, secondly utilizing the $V_{rad}(\varphi)-I_C(\varphi)$ relation that was shown to be valid for RRab stars in \cite{cikk}. For variables showing no or only small amplitude light curve modulation the two types of initial $V_{rad}(\varphi)$ functions led to similar final results. The $A_{V_{rad}} - V_{Amp}$  relation of individual Blazhko variables during the modulation cycle is, however, steeper than that was defined by \cite{liu}, and from the $V_{rad}(\varphi)-I_C(\varphi)$ relation valid for nonmodulated RRab stars \citep[see Fig. 2 in][]{dist}. Blazhko variables obey the same amplitude relation as normal RRab stars at around their maximum amplitude phase \citep{dist}. As a consequence, the amplitude of the mean radial velocity variation of a Blazhko star is supposed to be smaller than that of a nonmodulated star  with photometric amplitude similar to the amplitude of its mean light curve. The slope of the $\Delta A_{V_{rad}}/\Delta A(V)$ amplitude ratio is 30 for regular RRab stars while it is about 41 for Blazhko variables. The amplitudes of the initial radial velocity curves used in the analysis of MW Lyr, calculated either from the $I_C$  light curves or Liu's template, are therefore scaled differently from normal RRab stars. Both mean light curve sets are modelled assuming three values for the $\Delta A_{V_{rad}}/\Delta V_{Amp}$ ratio, 37, 41, and 45, respectively, and with zero points of the  $A_{V_{rad}} - V_{Amp}$ relation determined from matching the maximum amplitude value of the $V_{rad}$ curve to the  $A_{V_{rad}} - V_{Amp}$ relation of nonmdulated RR Lyrae stars.

\begin{table}
\caption{Global mean physical parameters of MW Lyr }
 \label{tab1}
\begin{tabular}{rrrrrrr}
\hline
d [pc] & $\overline{M_V}$ & $\overline{L/L_{\odot}}$& $\overline{T_{eff}}$ & $\overline{R/R_{\odot}}$ & $\overline {\log g_{stat}}$ &$\overline{{\mathfrak{M/M}}_{\odot}}$\\

\multicolumn{3}{l}{uncertainties}&&&&\\
\hline\hline
\multicolumn{7}{c}{time transformed data}\\
\hline
\multicolumn{5}{l}{$\Delta A_{V_{rad}}/\Delta A(V) = 37$}&\\
\hline
3595    &   0.74 & 44.7 & 6892.5   &  4.64 &2.982    &0.76\\
147      &  0.09 & 3.6 & 2.5       & 0.19 & 0.010 & 0.08\\
\hline
\multicolumn{5}{l}{$\Delta A_{V_{rad}}/\Delta A(V) = 41$}&\\
\hline
3460   &    0.83 & 41.4 & 6891.9  &  4.47 & 2.973& 0.69\\
145     &   0.09 &  3.4 &2.5      & 0.18 & 0.010 & 0.07\\
\hline
\multicolumn{5}{l}{$\Delta A_{V_{rad}}/\Delta A(V) = 45$}&\\
\hline
3343   &    0.90  & 38.6& 6890.3     &  4.32 & 2.965& 0.63\\
138   &    0.09  & 3.2 &2.4  &         0.18 & 0.010 & 0.06\\
\hline\hline
\multicolumn{7}{c}{original data} \\
\hline
\multicolumn{5}{l}{$\Delta A_{V_{rad}}/\Delta A(V) = 37$}&\\
\hline
3367    &   0.89 &  39.0 &  6889.4 &       4.35 & 2.966  & 0.64\\
106     &   0.07 &  2.4 &  2.7 &            0.14 &  0.009 &  0.05\\
 \hline
\multicolumn{5}{l}{$\Delta A_{V_{rad}}/\Delta A(V) = 41$}&\\
\hline
3223     &  0.98 &35.7 &  6888.0 &       4.16 &2.956    & 0.57\\
111       & 0.08 & 2.4 &  2.3 &         0.14 &  0.008 &  0.05\\
\hline
\multicolumn{5}{l}{$\Delta A_{V_{rad}}/\Delta A(V) = 45$}&\\
\hline
3104     &  1.06& 33.2  &  6887.0 &       4.01  &2.947 &  0.52\\
109      &  0.08 & 2.3 &  2.3 &          0.14 &  0.009 &  0.05\\
\hline
\end{tabular}
\end{table}

As there is no information on the changes of the actual shapes of the $V_{rad}(\varphi)$ curves during the Blazhko cycle we do not know whether any of the two approximations of the radial velocity variation that the IP method uses is valid for Blazhko variables. In the first method when using Liu's template, the shape of the initial $V_{rad}$ curve is always the same only its amplitude is scaled, while in the second one when calculating the $V_{rad}(\varphi)$ curve from the actual $I_C$ light curve, it varies in accordance with the variations of the light curve's shape. Having no a priori knowledge on the $V_{rad}(\varphi)$ of a Blazhko star, in order to make even less restrictions on its shape, in the IP process an even smaller weight of the initial $V_{rad}(\varphi)$ curve is given than that has been used for unmodulated RRab stars in \cite{cikk} to allow a larger variance of its Fourier parameters.

Table~\ref{tab1} summarizes the mean global parameters of MW Lyr using the original and the time transformed data and 3 possible values for the $A_{V_{rad}} - V_{Amp}$ relation of Blazhko variables. The estimated uncertainties are the rms scatter of the results running the IP method using 8 settings:

\begin{enumerate}
\item {Initial $V_{rad}(\varphi)$ calculated from Liu's $V_{rad}$ template curve or from the $V_{rad}(\varphi)-I_C(\varphi)$ relation.}
\item {Large and small weights are given to the initial  $V_{rad}(\varphi)$  function, the first keeps it close to its initial shape, the second allows the amplitudes and phases of the lower Fourier components of the $V_{rad}(\varphi)$ curve to vary substantially. In \cite{cikk} it was shown that giving small weights to the initial  $V_{rad}(\varphi)$ curve leads to unreliable shape of the solution  $V_{rad}(\varphi)$ curve. The results of the IP method are the most reliable for unmodulated RRab stars if large and medium weights of the initial  $V_{rad}(\varphi)$ curve are given. However, as we do not know how the radial velocity curve of a Blazhko variable vary, we decided to use small weights of the initial  $V_{rad}(\varphi)$ curves instead of medium weights. Any result that is independent from this choice is practically independent of the exact shape of the radial velocity variations, and as so, is a robust solution for the detected  changes during the Blazhko cycle.}
\item {Input data are the Fourier fits of the $V$, $B-V$, and $V-I_C$ or the $V, B,$ and $I_C$ time series. These input data are not identical representations of the observations because the differences between the Fourier fits of the magnitudes do not reproduce exactly the Fourier fits of the colour indices.}
\end{enumerate}

As mentioned earlier and discussed in Paper I, we regard the time transformed data that corrects the phases of the pulsation to eliminate the phase modulation/pulsation period changes to be a more reliable representation of the data than the original time series. Therefore, in the detailed analysis of the light curves in different phases of the modulation, the distance and mass of MW Lyrae are fixed to the mean values of the results obtained for the time transformed data assuming $\Delta A_{V_{rad}} / \Delta V_{Amp} = 41 $, namely to 3460 pc  and 0.69 $\mathfrak{M_{\odot}}$ values, respectively. This mass is somewhat larger than the mass of higher metallicity horizontal branch stars in the instability strip according to stellar evolutionary models (e.g., ACS Survey, \citealt{dotter}; Padova evolutionary database, \citealt{salasnich}; Y2 evolutionary tracks, \citealt{demarque}). The IP method calculates the mass of the star from the pulsation equation of fundamental mode variables. The masses of some of the unmodulated RRab stars derived by the IP method are too large on evolutionary grounds, but the combination of the results of direct Baade-Wesselink analysis with the pulsation mass formula may also lead to too large mass values compared to the evolutionary values in some cases. E.g., 0.87 and 0.85 $\mathfrak{M_{\odot}}$ masses were derived in \cite{cikk}  for WY Ant and UU Cet, respectively, from the pulsation equation and the results of the Baade-Wesselink analysis of \cite{kbw}. To resolve this discrepancy is, however, beyond the scope of the present paper.

It is important to note, however, that when running the IP code using any of the physical parameter combinations listed in Table 1, similar solutions for the variations of the global physical parameters with the Blazhko phase are obtained. It is found that only the mean values of these parameters depend on which  mean light curve solution is accepted. All the conclusions of the next sections remain unchanged if other values of the fixed parameters, [Fe/H], distance and mass, were selected, and analyses are performed using the original data with other possible values of the  $\Delta A_{V_{rad}} / \Delta V_{Amp}$ amplitude ratio.

\begin{figure}
  \includegraphics[width=9.3cm]{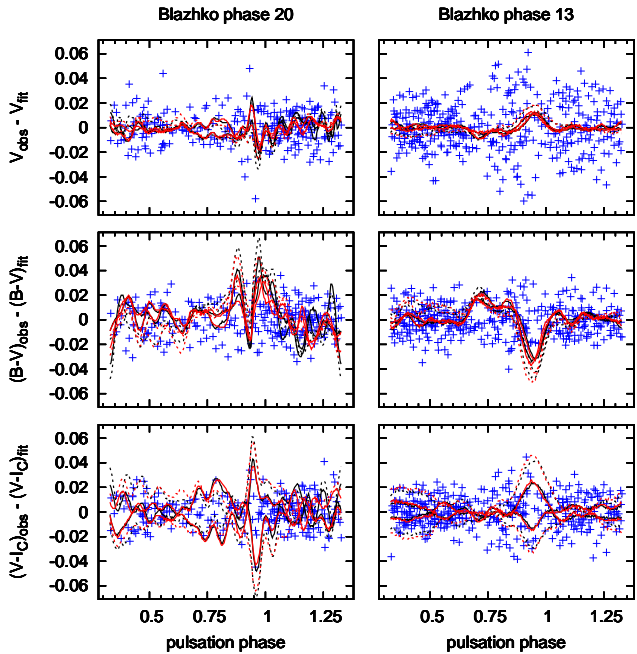}
  \caption{Residual $V$, $B-V$, and $V-I_{C}$ data (crosses) are plotted for the largest and smallest amplitude phases of the modulation, for Blazhko phases 20 and 13, respectively. The mean pulsation curves in both phases are subtracted. The lines show the deviations of 8 variant fits of the IP method from the mean pulsation curves.}
\label{res}
\end{figure}

\section{Changes in the physical parameters during the Blazhko cycle}
The detailed analysis of the light curves in different phases of the modulation is carried out by running the IP code using atmosphere models with [Fe/H]=$-0.5$ on 20 segments of the time transformed photometric data with the 8 settings specified in (i), (ii), and (iii) points.  The distance and mass of MW Lyr are fixed to the 3460 pc and 0.69 $\mathfrak{M_{\odot}}$ values, respectively.

The IP method fits the light and colour curves by finding the appropriate changes in the surface radius and temperature of the star. The fitting accuracy is about 0.02 mag. Just for comparison, in Paper I we have shown that due to some stochastic/chaotic behaviour of the modulation the best Fourier fit to the complete $V$ data set reproduces the observations with only 0.02 mag accuracy. The Fourier fits of the light curves of the time transformed data in the 20 phase bins of the modulation give, on average, also 0.02 mag rms residual, which is significantly larger than the observational inaccuracy. This is partially due to modulations detected in Paper I with periods that are non-integer multiplets of the main modulation period, and to small but nonperiodic perturbations of the modulation. Fig.~\ref{res} plots the residual $V, B-V$, and $V-I_{C}$ data in the largest and smallest amplitude phases of the modulation, in Blazhko phase 20 and 13, respectively. The deviations of the fitted curves from the mean light curves of the data are also drawn. The fits seem to match the light curves within the required accuracy, with the largest deviations at around the rising branch of the light curves (phase 0.85-1.00) where the observed colours of the dynamic atmosphere of RR Lyrae stars definitely cannot be fitted accurately  with static atmosphere model results. Moreover, in this phase of the pulsation the modulation is not strictly 
regular (as it was shown in paper I), which may also explain the larger 
residuals of the solutions of the IP method here.

\begin{figure}
  \includegraphics[width=9.3cm]{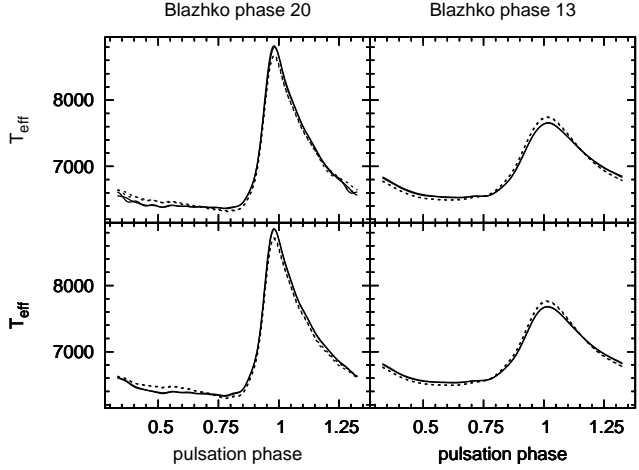}
  \caption{$T_{eff}$ variations in the large amplitude (left panels) and small amplitude (right panels) phases of the modulation. Top and bottom panels show results if the initial $V_{rad}$ curve is defined from the $V_{rad}-I_C$ relation and from the template curve given by \citet{liu}, respectively.  Four curves are plotted in each panel, giving large (solid lines) and small weights (dashed lines) to the initial $V_{rad}$ curves and using the $V, B-V$, and $V-I_C$ or the $B,V,$ and $I_C $ curves as input data. These latter curves, however, highly overlap in each of the plots shown in Figs.~\ref{t}-\ref{lg}.}
\label{t}
\end{figure}

\begin{figure}
  \includegraphics[width=9.3cm]{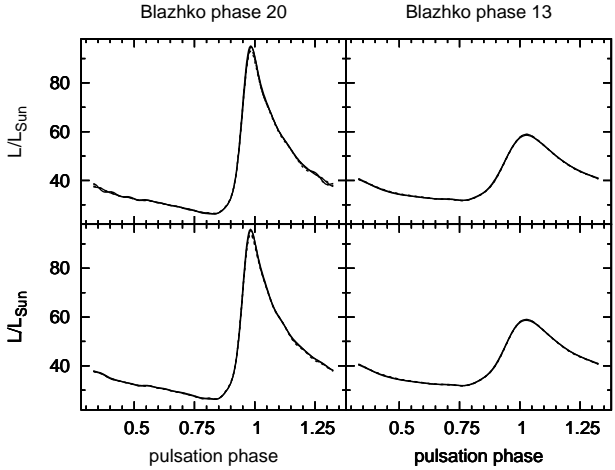}
  \caption{The same as Fig.~\ref{t} but for the luminosity variations. }
 \label{l}
\end{figure}

\begin{figure}
  \includegraphics[width=9.3cm]{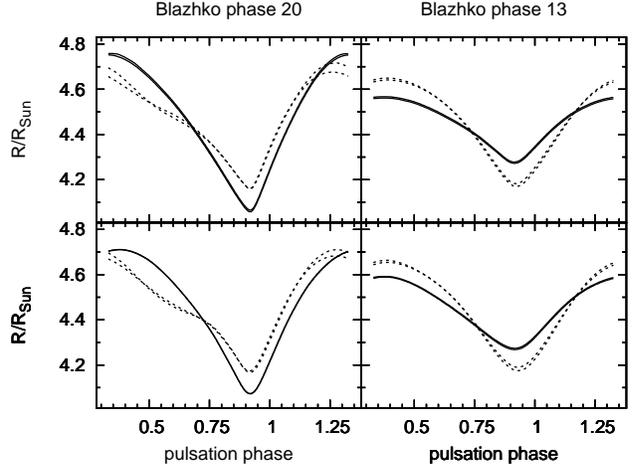}
  \caption{The same as Fig.~\ref{t} but for the radius variations.}
\label{r}
\end{figure}

\begin{figure}
  \includegraphics[width=9.3cm]{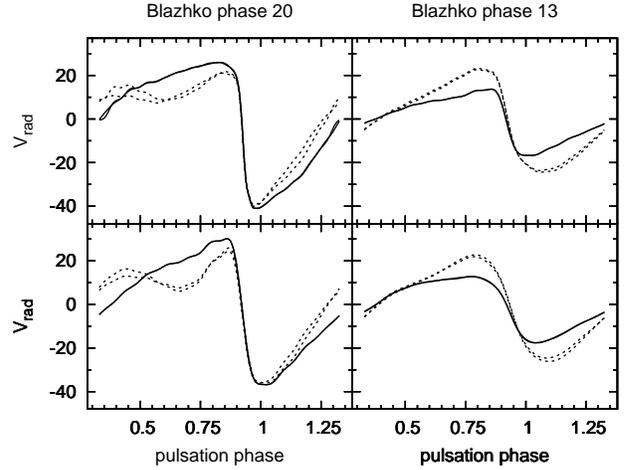}
  \caption{The same as Fig.~\ref{t} but for the radial velocity curves.}
\label{vr}
\end{figure}

\begin{figure}
  \includegraphics[width=9.3cm]{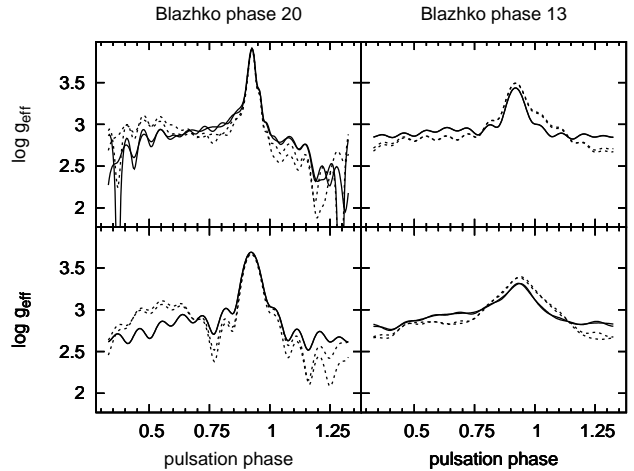}
  \caption{The same as Fig.~\ref{t} but for the effective surface gravity variations.}
\label{lg}
\end{figure}

In Figs.~\ref{t}-\ref{lg} the results of the IP method for the ${T_{eff}}, L/L_{\odot}, R/R_{\odot}, V_{rad},$ and  $ \log g_{eff}$  changes are shown in two phases of the Blazhko modulation, one at the largest and another at the smallest pulsation amplitude. The upper panels show when the initial $V_{rad}$ curve is calculated according to the $V_{rad}-I_C$ relation while bottom panels show results using the template  $V_{rad}$ curve defined by \cite{liu}. Solid lines show the results when the $V_{rad}$ curve is kept close to its initial shape, while the dashed lines indicate how the results change if the lower Fourier parameters of the initial $V_{rad}$ curves are allowed to vary. Each setting is applied using either the fits to the $B, V,$ and $I_C$ or the $V, B-V $and $V-I_C$ data, however, these differences have only very marginal effect on the results.

\begin{figure}{\begin{center}
  \includegraphics[width=6.cm]{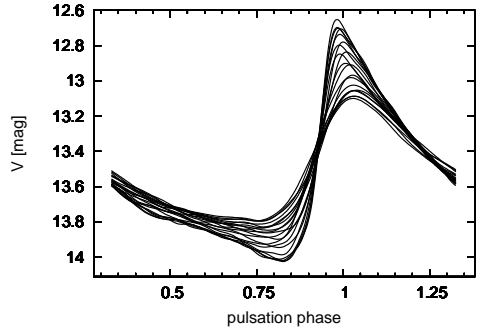}
\end{center}}
  \caption{The time transformed $V$ light curves of MW Lyrae in 20 phase bins of the modulation.}
\label{mindenv}
\end{figure}

\begin{figure}
  \includegraphics[width=9cm]{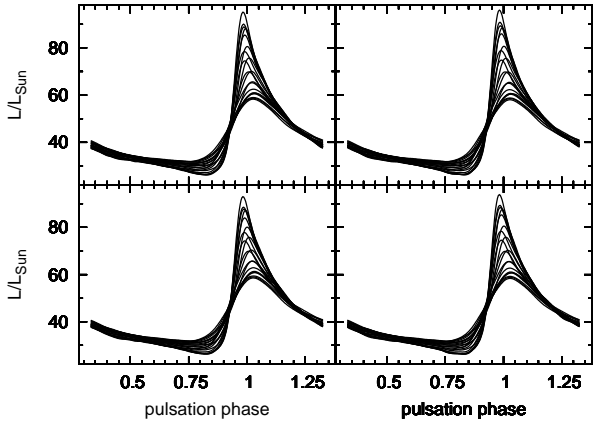}
  \caption{Luminosity variations in 20 phase bins of the modulation. Left and right panels show results using the transformation of the $I_C$ light curves and Liu's template as initial radial velocity curves, respectively. In the top panels the $V_{rad}$ curves are kept close to their initial curve, while in the bottom panels the initial $V_{rad}$ curves ar allowed to vary significantly in the fitting process.}  
\label{mindenl}
\end{figure}

\begin{figure}
  \includegraphics[width=9cm]{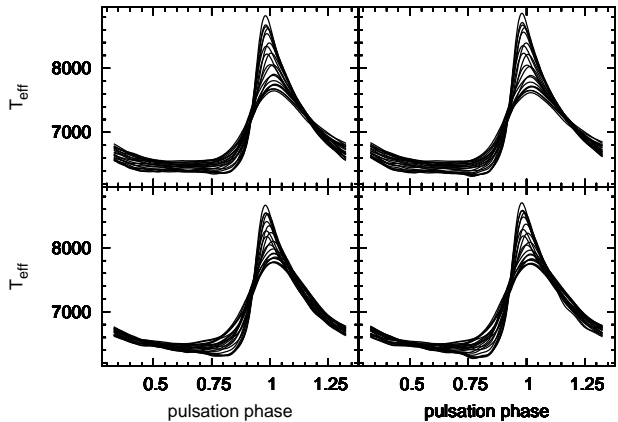}
  \caption{The same as in Fig~\ref{mindenl} but for the temperature variations in 20 phase bins of the modulation.}
\label{mindent}
\end{figure}

\begin{figure}
  \includegraphics[width=9cm]{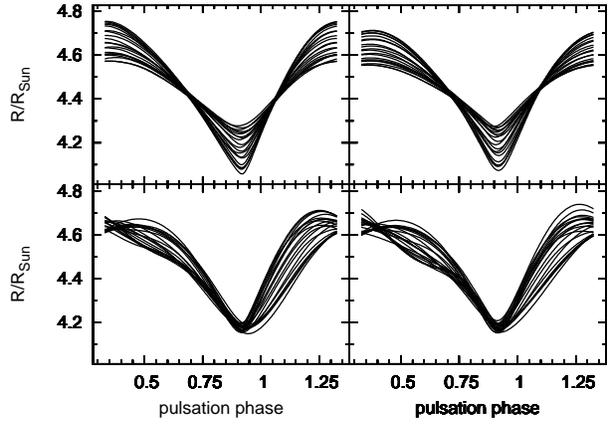}
  \caption{The same as in Fig~\ref{mindenl} but for the radius variations in 20 phase bins of the modulation.}
\label{mindenr}
\end{figure}

\begin{figure}
  \includegraphics[width=9cm]{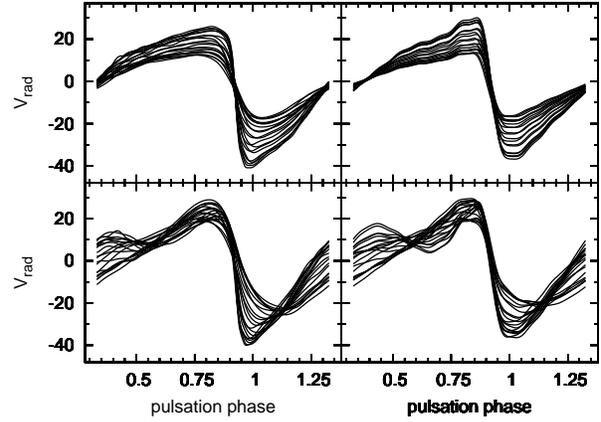}
  \caption{The same as in Fig~\ref{mindenl} but for the radial velocity curves in 20 phase bins of the modulation.}
\label{mindenvr}
\end{figure}

\begin{figure}
  \includegraphics[width=9cm]{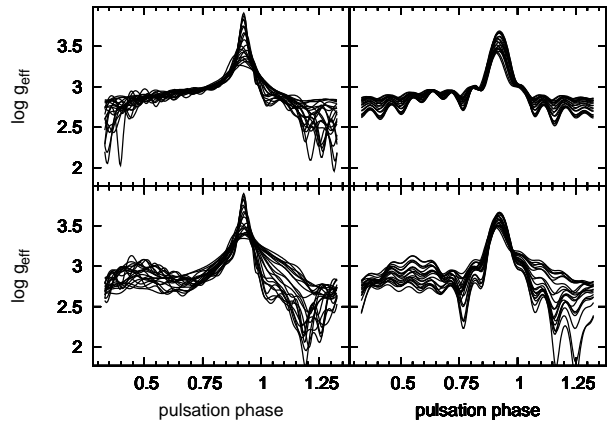}
  \caption{The same as in Fig~\ref{mindenl} but for the effective surface gravity variations in 20 phase bins of the modulation.}
\label{mindeng}
\end{figure}

Fig.~\ref{mindenv} displays the light curve variation of MW Lyr in 20 phase bins of the modulation according to the time transformed data. For comparison, in Figs.~\ref{mindenl}-\ref{mindeng} the variations of the physical parameters derived form the IP method are shown. The left and right panels show if the initial $V_{rad}$ curves are calculated from the $I_C$ light curves or from Liu's template, while top and bottom panels are for large and small weights of the initial $V_{rad}$ curves, respectively.

There are only minor changes in the luminosity and surface effective temperature curves depending on the choice and weights of the initial $V_{rad}$ curves. The amplitude of the temperature and luminosity variations during the pulsation of the star are significantly different in the different phases of the modulation. The temperature and luminosity at maximum amplitude vary between 6300 and 8800 K and between 26 and 95 L$_\odot$, while at minimum amplitude only between 6500 and 7700  K, and between 31 and 58  L$_\odot$, respectively. The changes in the amplitude of the temperature and luminosity variations are about $50 \%$!  These results are very robust, the ranges of the detected changes in the variations of these parameters in different phases of the modulation are independent of the choice and alterations of the initial radial velocity curve.

Substantial changes in the radius, radial velocity and surface effective gravity curves can be also seen. It is important to note that in those solutions when the shape of the initial radial velocity curve is allowed to vary substantially the results are very similar if  Liu's template is used or the initial $V_{rad}$ curve is calculated from the $I_C$ light curve. The radial velocity and, as a consequence, the radius variations are somewhat different if the $V_{rad}$ curve is kept close to its initial template, or it is allowed to vary significantly. In the first case significant changes in the amplitude of the radius variation during the Blazhko cycle can be detected, while in the second case not the amplitude but the shape of the radius curves vary. The only radial velocity observations of a Blazhko variable with good enough phase coverage of both the pulsation and the modulation cycles were recently obtained for RR Lyr \citep{chadid}. Fig.~9 in \cite{chadid} shows that the radial velocity curves of RR Lyr remain smooth during the modulation and there is no sign of a double wave shape similar to what our results indicate if the shapes of the radial velocity curves are allowed to differ significantly from their initial templates during the large amplitude phase of the modulation. Most probably the real variations of the radial velocity curves are close to those shown in the upper panels of Fig.~\ref{mindenvr} and not those in the bottom panels, i.e, the radial velocity curves of Blazhko variables differ from the $V_{rad}$  templates of unmodulated RRab stars mostly in their amplitudes and not in their shapes. If this is indeed the case, then there is also about $50\%$ change in the amplitude of the radius variation of the pulsation during the Blazhko cycle.
\vskip 1cm

\section{Changes in the mean physical parameters during the Blazhko cycle}

From photometric data different types of mean magnitudes and colours can be derived: average values of the intensities or the magnitudes or mean colours calculated  from the difference of the mean magnitudes or from colour data. The intensity averaged magnitudes are brighter than the magnitude averaged ones, while for the colours the $(m_1-m_2)> \langle m_1-m_2 \rangle > \langle m_1 \rangle - \langle m_2 \rangle$  relation holds. It was shown, however, by \cite{bono} that none of these mean magnitudes and colours match the equivalent static values of RR Lyrae stars in the full ranges of their possible parameter domain. Therefore, to draw any conclusion for the changes of the mean physical parameters during the Blazhko cycle from the changes of the mean colours and magnitudes is somewhat ambiguous.

\begin{figure*}
 \includegraphics[width=16.6cm]{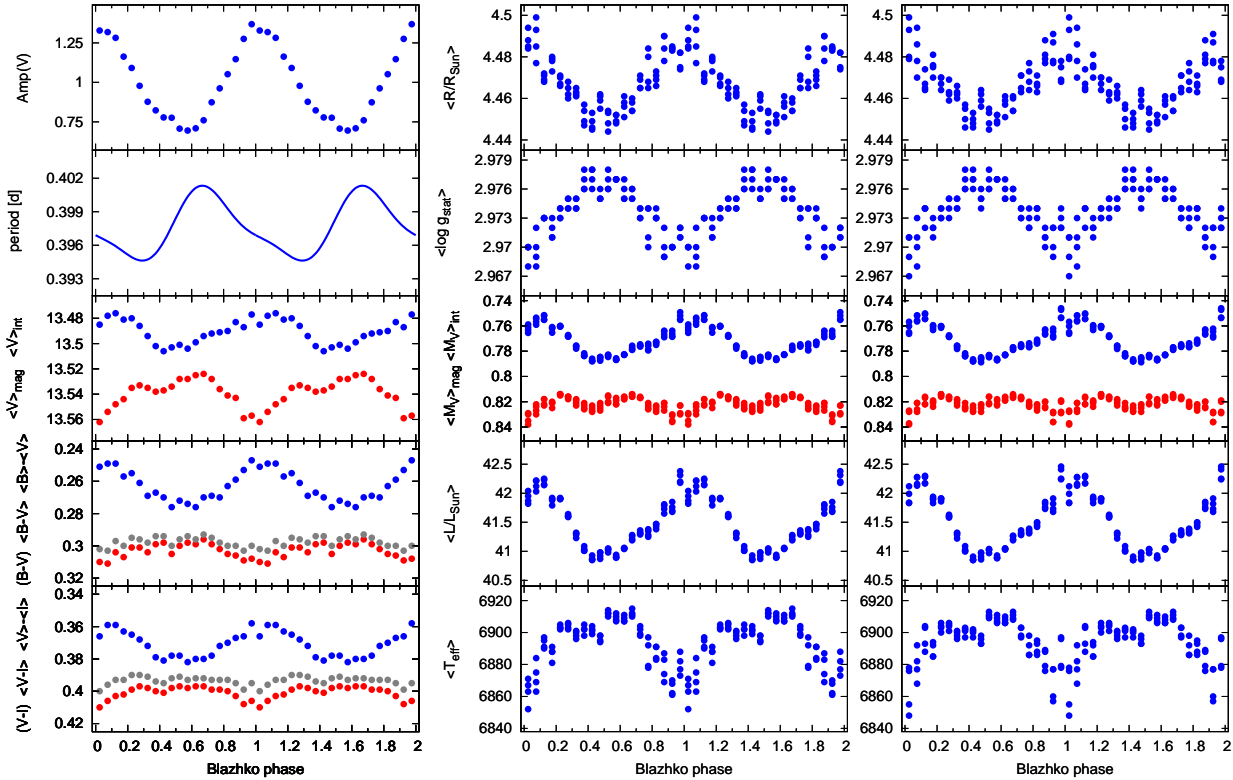}
  \caption{Observed mean parameters (Amp(V), pulsation period, intensity and magnitude averaged $V$ brightnesses, and three types of average values of the $B-V$ and $V-I_C$ colours) are shown in the left panels. Middle and right panels show the variations of calculated mean parameters if initial $V_{rad}$ curves are defined according to the $V_{rad}-I_C$ relation, and Liu's template $V_{rad}$ curve, respectively. For each phase bins the values of 4 solutions with different settings are shown to illustrate the inherent uncertainty the the results.}
\label{minden}
\end{figure*}

In the left panels in Fig.~\ref{minden} the observed mean data derived from the photometry of MW Lyr are shown for 20 Blazhko phase bins. The period changes correspond to the pulsation period variation determined in Paper I from the phase of the $f_0$ pulsation frequency. These plots show definitely that each of the mean colours and magnitudes vary with 0.01-0.02 mag amplitude, but these changes are of the opposite sign for the different averages, making any firm conclusion of the mean temperature and luminosity changes impossible.

However, after finding the appropriate pulsation curves of the physical parameters in different phases of the modulation with the IP method we can derive their mean values directly. In the middle and right panels in Fig.~\ref{minden} the mean physical parameter changes are shown if the  $V_{rad}-I_C$ relation and Liu's template are used to define the initial $V_{rad}$ curves, respectively. For each phase bin four data of the mean parameters are plotted corresponding to the results with different settings of the IP method as shown in the panels in Figs.~\ref{t}-\ref{lg}.

Fig.~\ref{minden} shows unambiguously that during the Blazhko cycle the mean physical parameters do indeed vary. The amplitudes and signs of these variations are independent of the choice of the initial $V_{rad}$ curve and of which setting of the IP code is used. We have also tested whether there is any change in these variations if any other possible combinations of the mass and distance of MW Lyr as listed in Table~\ref{tab1} are adopted, or if the original data are used instead of the time transformed ones, or if different $\Delta A_{V_{rad}}/\Delta A(V)$ ratio is applied.  Any of these possibilities have only minor influence on the results shown in Fig.~\ref{minden}, only the parameters' ranges are shifted corresponding to the data given in Table~\ref{tab1}.

We can therefore sate that, for the first time, we succeeded in reliably detecting changes in the mean global physical parameters of a Blazhko variable during the Blazhko cycle. $1-2\%$ changes in each of the parameters are evident. MW Lyr is larger by about $0.04 R_\odot$, more luminous by about $1.00 L_\odot$ and cooler by about 50 K in the large pulsation amplitude phases of the modulation as compared to its small pulsation amplitude phases. Though we cannot exclude the possibility that the detected changes in the mean global parameters arise from a complicated conjunction of different unknown processes, the most straightforward conclusion is that they simply reflect real changes in the mean global physical parameters. 

The relative radius variation $\Delta R / R$ is 0.009 which leads to  $\Delta P / P =0.014$ relative period change as a consequence of the $\Delta P / P \approx 3/2 \times \Delta R / R$ pulsation relation. The detected changes in the pulsation period is  $\Delta P / P= 0.016$. This good agreement between the period changes  measured directly from the phase modulation and calculated from the radius variation proves that both the measured radius changes and our treatment to regard the phase modulation of the pulsation as pulsation period changes should be real. The phase relation between the radius and period changes are not synchronized, the period is about the longest when the mean stellar radius and the pulsation amplitude is the  smallest. \cite{stothers} found a similar phase connection between the pulsation amplitude and pulsation period changes in RR Lyrae. 

Our results also reveal that the intensity mean $V$ magnitude reflects the luminosity changes  and the $(B-V)$ and $\langle B-V \rangle$, or $(V-I_C)$ and $\langle V-I_C \rangle$ mean colours reproduce the temperature changes correctly, notwithstanding the large amplitude variation of the pulsation. Consequently, these values can be regarded as good representatives of the  equivalent static values of RR Lyrae stars. The temperatures derived from $\langle B \rangle - \langle V \rangle$ or $\langle V \rangle - \langle I_C \rangle$ colours may differ from these temperatures by 100-250 K, the difference is the largest when the amplitude of the pulsation is large.  

\section{Comparison with pulsation model results}

Assuming that the variation of MW Lyr can be fully described by radial pulsation, some hints on the character of the variation of the mean physical parameters can be drawn from numerical pulsation calculations. Since there are no nonlinear pulsation models that can reproduce the Blazhko phenomenon, such calculations can reveal only partly the relations found in this paper. For our calculations we used the Florida-Budapest pulsation code with standard RR Lyrae parameters \citep[see][]{model}.
Even the state of the art pulsation codes contain only a relatively simple model of convection. In these models only one dynamical equation is used for the turbulent energy, with additional source functions of turbulent energy and convective flux. According to our experience, the interaction of this simple turbulent equation with the hydrodynamics and radiation transfer does not give rise to any time-scale or modulation that can be connected to Blazhko cycles. If turbulence plays an important role in Blazhko phenomenon, as suggested by \cite{stothers}, then to model the effect, additional physics, e.g. with a more complicated turbulent-convection formalism, or with the inclusion of interaction with magnetic field, should be included in the codes. In the lack of any more sophisticated model only some aspects of the nonlinear dynamics of pulsation can be tested in comparison with the observations.

The natural timescales that come from numerical calculations are the pulsation period and the e-fold time of amplitude variation during the onset of pulsation (the last one is related to the linear growth rate of pulsation). Since the growth rate normalized with the pulsation frequency, is of the order of a few percent, the e-fold time of pulsation growth or decay is of the order of the Blazhko period. It means that any change in the stellar structure can be realized in pulsational changes on the period of the Blazhko cycle. 

As a first test, nonlinear pulsation calculations are initiated with a small velocity perturbation with the corresponding eigenvector of the linear stability analysis. Then the onset of pulsation occurs through a transient, with the growth of pulsation amplitude. We used this phase of the numerical calculations to estimate the dependence of the mean physical quantities on the amplitude.

A more physical test is to check the effect of the variation of turbulent parameters on the mean physical parameters. From the 6 turbulent ($\alpha$) parameters of the Florida-Budapest code we selected the efficiency of eddy viscosity (the dissipation due to turbulence), since this parameter ($\alpha_\nu$) does not affect the static structure and the linear growth-rates of the pulsation modes. With this test the amplitude of the relaxed pulsation can be changed and it again gives the dependence of mean values of some of the global parameters on the amplitude. From a finite amplitude (limit cycle) model the hydrocode can be restarted with a slightly modified eddy viscosity parameter, which gives an efficient way to get the pulsation model for different values of $\alpha_\nu$, and finally the averaged physical quantities as the function of amplitude. 

\begin{figure}
  \includegraphics[width=8.9cm]{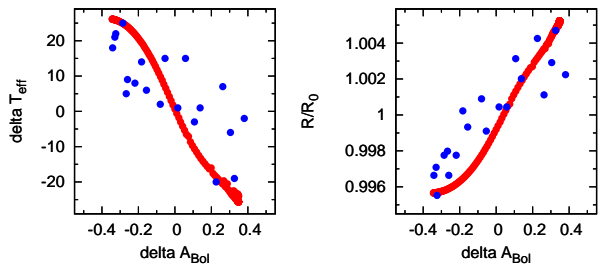}
  \caption{The observed changes of the mean temperature (left panel) and relative mean radius (right panel) variations as a function of the bolometric amplitude variation of MW Lyr in different phases of the modulation (dots) are compared with pulsation model predictions (lines).}
\label{zoli}
\end{figure}

The two tests gave basically identical results, proving that the amplitude dependence of some of the averaged
quantities is not a transient event, but a real dynamic effect. We calculated several models of both fundamental and overtone pulsations. As expected, the mean radius of the star increases with the pulsation amplitude according to the results of both tests. We found that approximately $\Delta R/R_0 \approx 0.015 \Delta A_{Bol}$ holds. The mean effective temperature decreases with increasing amplitude: $\Delta T_{eff} \approx 100 \cdot \Delta A_{Bol}$. These results agree well with the observed dependencies as shown in Fig~\ref{zoli}.  The figure compares the observed relative changes in mean temperature and radius as a function of the observed bolometric amplitude (calculated from the luminosity curve solution of the IP method) with model predictions. The measured amplitude dependencies are $\Delta T_{eff} \approx 80 \cdot \Delta A_{Bol}$ and $\Delta R/R_0 \approx 0.012 \Delta A_{Bol}$.

We have to note, however, that the calculations reveal only one of the mechanisms that should be taken into consideration to understand the Blazhko effect, namely the standard dynamics of radial pulsation. For example, if one considers the standard relation between $\Delta P/P$ and $\Delta R/R_0$, the order of the observed variations of these quantities agrees with each other as shown in Sect. 4. But since the whole mechanism is more complicated, it is not astonishing that the period and radius variations are not in phase as expected from the simple relation. Since in these tests the static structure of the star has been fixed, it is not expected that all the observed variations during the Blazhko cycle could be estimated. First of all, this simplified dynamic system has only a very limited effect on the pulsation period (of the order of  $10^{-4}$ ). It indicates that structural changes are necessary to reproduce all the observational data, e.g. by varying the other turbulent parameters in the code. 

Another deficiency of our simple model is that since there is no built in mechanism besides the kinetic energy of pulsation in the code, which results in time varying storage of radiative energy on a longer time scale than the pulsation, the mean luminosity changes cannot be compared to the observations. Moreover, dynamical effects can also play an important role in the complete picture of the phenomenon. This effect, however, cannot be investigated without a self-consistent model which naturally provides the periodic or quasi periodic modulation of turbulent structure. To get a consistent model of Blazhko phenomenon, based on the dynamics of turbulent convection, more sophisticated models and lot more computational efforts are needed. The goal of this paper cannot be to find the solution of the long standing problem of Blazhko effect. However, it turns out, that some of the observational results can be  well estimated even with our simple treatment. Whether it is a fortunate coincidence or the indication that variations on the turbulent convective structure of RR Lyrae stars are an important ingredient of Blazhko phenomenon, should be answered in further studies.

\section{Colour dependent characteristics of the modulation and pulsation frequencies}

The detected changes in the mean global physical parameters during the Blazhko cycle prove that the physically meaningful frequency of the modulation is $f_m$, the frequency of the modulation itself, and not any of the side lobe frequencies appearing close to the pulsation frequency or its harmonics in the Fourier spectrum of the light curve. A similar conclusion was already drawn in \cite{sscnc} based on the colour behaviour of the amplitudes and the phases of the pulsation, modulation side lobe, and the modulation frequencies of RR Gem and SS Cnc. It was found that while the modulation side lobe frequencies have similar amplitude ratios and phase differences in the different colours as the pulsation frequencies have, the modulation frequency has discrepant values.

Table~\ref{f0} summarizes the amplitude ratios of the pulsation and modulation frequencies of MW~Lyr calculated from the Fourier parameters given in Table 5 in Paper I. Only frequency components with $V$ amplitudes larger than the $V$ amplitude of the $f_\mathrm{m}$ frequency are considered. For MW Lyr the phase differences of the $f_\mathrm{m}$ frequency component do not show discrepant behaviour but its $A(B)/A(V)$ amplitude ratio is about 0.1 larger, and its $A(V)/A(I_C)$ amplitude ratio is 0.1 smaller than the amplitude ratios of the pulsation and modulation side lobe frequency components. 

Based on, on the one hand, the similarity between the colour dependence of the amplitudes of the modulation side lobe frequencies and that of the pulsation frequencies and, on the other hand, on the discrepant amplitudes of the modulation frequency itself, we conclude that the side lobe frequencies are  most probably combination frequencies of the pulsation frequencies ($kf_0$) and the modulation frequency ($f_\mathrm{m}$). As being combination frequencies they inherit the properties of their larger amplitude component, namely that of a pulsation frequency component. In this case, the independent frequency of the modulation is, in fact, the modulation frequency, $f_\mathrm{m}$.

\begin{table}
\caption{Comparison of the colour behaviour of the pulsation and modulation side componets with the properties of $f_\mathrm{m}$ }
 \label{f0}
\begin{tabular}{lc@{\hspace{-1pt}}c@{\hspace{-1pt}}c@{\hspace{-1pt}}c}
\hline
frequencies& $A(B)/A(V)$& range&$A(V)/A(I_C)$& range\\
 \hline
$f_\mathrm{m}$&1.379&&1.400&\\
 \hline
$kf_0\,{^a}$                 &$1.263$&$1.185 / 1.336$& $1.522$&$1.438 / 1.625$\\
$kf_0\pm f_\mathrm{m}\,{^b}$ &$1.302$&$1.251 / 1.352$& $1.577$&$1.505 / 1.623$\\
\hline
\multicolumn{5}{l}{\scriptsize{${^a}$ average value for 6 components}}\\
\multicolumn{5}{l}{\scriptsize{${^b}$ average value for 10 components}}\\
\hline
\end{tabular}
\end{table}

\section{Conclusions}

As a conclusion, based on the results shown in Paper I and in this paper we can interpret the Blazhko modulation of MW Lyr as follows.

\begin{itemize}
\item
{The frequency that characterizes the Blazhko modulation of RRab stars is the modulation frequency,  $f_\mathrm{m}$, as we have found that the observed modulations are governed by global physical changes that also occur with the Blazhko period. If this proves to be indeed the case, then theories that tie the Blazhko modulation to a side frequency of a pulsation frequency component ($kf_0\pm f_\mathrm{m}$) misinterpret the Blazhko phenomenon \citep[e.g.][]{dz}.}
\item
{The global mean physical parameters of the star change about $1-2\%$ during the Blazhko cycle. Most probably the primary changes occur in the mean stellar luminosity and the detected changes in mean stellar temperature, radius, and pulsation period are the consequences of the luminosity changes. }
\item
{The phase modulation during the Blazhko cycle reflects simply the oscillations of the pulsation period, while the amplitude changes are due to the changes in the mean global physical parameters of the star (L, T, R) reflecting periodic alterations in the atmosphere structure of the star.}
\end{itemize}

Numerical models of the onset of radial nonlinear pulsation display very similar dependence of mean radius and effective temperature on amplitude to the observed ones. It also suggests that, at least in part, the variation of mean quantities can be understood by the simple dynamics of radial pulsation. However, to unfold the whole picture, a presently unknown mechanism of the variation of the internal structure of the star (e.g., turbulent/convective properties) needs to be revealed.

Similar conclusions has been recently drawn about the nature of the Blazhko phenomenon by \cite{stothers}. He supposed that the underlying mechanism disturbing periodically the amplitude and period of the purely radial mode pulsation is cyclic weakening and strengthening of turbulent convection in the stellar envelope. The suggested triggering mechanism, namely a turbulent convective dynamo, would, however, most probably result in much less regular modulation behaviour than observed in MW Lyr. Another difficulty with Stother's idea is how the multiperiodic nature of the modulation observed in some Blazhko stars e.g., CZ Lac \citep[][and S\'odor, Jurcsik et al. in preparation]{sodor} and XZ Cyg \citep{xzc} can be explained. Consequently, while we think that though our results support the idea that during the Blazhko cycle the pulsation remains purely radial, further theoretical efforts are needed in order to find the reason for the drastic changes observed in its main properties.

\section*{Acknowledgments}

The authors would like to thank the comments of the anonymous referee which helped to improve the lucidity of the paper significantly.
The financial support of OTKA grants K-68626 and T-048961 is acknowledged. HAS thanks the US National Science Foundation for support under grants AST 0440061 and AST 0607249.

\end{document}